\documentclass{aa}  

\usepackage{graphicx}
\usepackage{capt-of}
\usepackage{txfonts}
\usepackage{lipsum}
\usepackage{float}
\usepackage{caption}
\usepackage{subcaption}         
\usepackage{lscape}             
\usepackage{placeins}           
\usepackage{color}
\usepackage{xcolor}

\usepackage{soul}
\definecolor{darkorange}{rgb}{1.0, 0.55, 0.0}

\begin{document}

   \title{Testing various assumptions for radiolysis, non-diffusive chemistry, and chemical desorption in cold cores}


%

   \author{V. Wakelam\inst{1} 
        \and T. Tu\inst{2}
        }

   \institute{Laboratoire d’astrophysique de Bordeaux, Univ. Bordeaux, CNRS, B18N, allée Geoffroy Saint-Hilaire, 33615 Pessac, France\\
             \email{valentine.wakelam@u-bordeaux.fr}
           \and School of Astronomy \& Space Science, Nanjing University, 163 Xianlin Avenue, Nanjing 210023, China}

   \date{Received xxx}

 
  \abstract
   {To study interstellar gas and grain chemistry, gas-phase astrochemical models have been developed since the early 1980s and gas-grain models since the 1990s. Each published model includes various assumptions mainly for surface processes.}
   {In this paper, we compare recently added mechanisms for grain surface chemistry, namely, non-diffusive chemistry, radiolysis, and chemical desorption.} 
   {Several formalisms for these processes have been added to our astrochemical model Nautilus, and we tested them, comparing the predicted gas-phase and ice abundances. Our predictions are also compared to gas and ice observed compositions.}
   {Our main findings are that radiolysis itself does not influence the results. Non-diffusive chemistry can have an impact on the gas-phase and ice species, but it depends on the adopted formalism. In particular,  the   one of Shingledecker \& Herbst (2018) changes the main reservoirs of the species in the ices, impacting the species in the gas-phase as well. The adopted formalism for chemical desorption can produce differences in the gas-phase by up to a factor of ten. Last, our standard model, without non-diffusive chemistry and with the chemical desorption from Fredon et al. (2021), produces the best results in relation to observed gas-phase abundances, while the ice observed agreement is unchanged. }
  {The formalism for some grain surface processes are important even for gas-phase abundances. More experiments are needed to constrain their efficiency, however. For the chemical desorption, each formalism relies on an uncertain parameter, which is the fraction of the energy actually delivered to the products, that can be adjusted to reproduce the experiments. }


   \maketitle
\nolinenumbers

\section{Introduction}

Carbon and heavier atoms are born in the centers of stars. At the end of their lives, stars spread their inner material into the interstellar medium (ISM), yielding at the same time the formation of interstellar dust refractory cores. In the diffuse medium, photo-dissociations by UV photons prevent the formation of molecules. Within large interstellar filaments, the material is organized as clouds (or clumps) of denser material. 

Numerical models have been developed since the early 1980s to study interstellar gas and grain chemistry\citep{1980ApJS...43....1P,1982ApJS...48..321G,1984ApJS...56..231L,1992ApJS...82..167H,1993MNRAS.263..589H}. If the gas-phase chemistry has been actively studied and improved over the past fifteen years thanks to a systematic review of gas-phase reactions \citep[see][and references therein]{2015ApJS..217...20W}, there is considerable room for improvement on the surface chemistry. Many modern astrochemical models have also been developed, suh as Nautilus  \citep{2016MNRAS.459.3756R}, MAGICKAL \citep{2011ApJ...735...15G}, GRAINOBLE \citep{2012A&A...538A..42T}, ROKKO \citep{2015A&A...584A.124F}, MONACO \citep{2017ApJ...842...33V},  and UCLCHEM \citep{2017AJ....154...38H}. Each of these models has different assumptions for grain surface chemistry, and they are constantly upgraded with new features. For example, Nautilus includes the desorption by sputtering induced by cosmic-rays, based on experimental results, while the other models do not. In MAGICKAL, \citet{2022ApJS..259....1G} includes a very large number of new processes. In ROKKO, the distribution of binding energy has recently been added \citep{2024ApJ...974..115F}, which is used instead of relying on one single value. Except for Nautilus and UCLCHEM, the codes are not public, so it is difficult to assess the differences in their results. In \citet{2025A&A...695A.247J}, we compared the ice composition obtained with some of these models in order to understand their prediction differences. We found major differences in all the ice species (H$_2$O, CO, CO$_2$, CH$_3$OH, NH$_3$, and CH$_4$). Again, given the complexity of these models, it was difficult to assess their origin, so some surface chemical pathways or the inclusion of non-diffusive chemistry were invoked.  

Three major model modifications have occurred in the recent years. The chemical desorption has been found to be highly efficient at medium densities \citep[about $10^4$ cm$^{-3}$;][]{2021A&A...652A..63W}. In this process, the energy released during an exothermic reaction at the surface of the grains can break the bonds of the products to the surface, resulting in their desorption  \citep{2007A&A...467.1103G,2016A&A...585A..24M,2016A&A...585A..55C,2018NatAs...2..228O}. The fraction of desorbed molecules depends on the exothermicity of the reaction, the binding of the species to the surface, and the fraction of the energy remaining in the product. This last parameter is difficult to assess because of a lack of experimental results.  The second modification is the new process of radiolysis, which is the chemical effect of cosmic rays in the ices. When a cosmic-ray particle collides with a grain, it can produce the dissociation of molecules and the production of suprathermal species, which diffuse at high velocity, boosting the chemical reactions \citep{2018PCCP...20.5359S,2018ApJ...861...20S,2019ApJ...876..140S,2020ApJ...888...52S}. The third modification is non-diffusive chemistry. This process has been introduced to mimic the fact that two species produced close to each other on the surface can react without diffusing \citep{2020ApJS..249...26J,2018PCCP...20.5359S,2025ApJ...990..163B}.

Our goal with the work presented in this paper was to test the efficiency of these new surface processes using Nautilus, as the processes are treated differently in various models. The paper is organized as follows. The model is presented in Sect. \ref{model-presentation} together with the different processes to test. The results of the different tests we made are given in Sect. \ref{results-section}. The different models are compared to gas and ice observations in Sect. \ref{obs-section}, while we provide our conclusions in the final section. 

\section{Model presentation}\label{model-presentation}

\subsection{General presentation}

Nautilus is a three-phase gas-grain astrochemical model. It computes the molecular and atomic abundances of gas and ice as a function of time starting from an initial chemical composition. It has three phases, which means that in addition to the gas, it computes the chemistry at the surface (the four most external monolayers) of the grains as well as in the bulk (the rest of the ice below the surface). The diffusion at the surface is more efficient, so the species are more reactive than in the bulk. Both the species on the surface and the bulk are dissociated by direct and indirect UV photons. Compared to previous versions of Nautilus, the photo-rates in the ices have been updated with the latest values in the gas-phase from kida.uva.2014 \citep{2015ApJS..217...20W}. The gas-phase network is kida.uva.2024 \citep{2024A&A...689A..63W}. 

The grain surface and bulk chemistry are computed using the rate equation approach. The species from the gas-phase remain at the surface of the grains, so they can diffuse and react upon encounters. Species from the surface can desorb due to temperatures breaking the bound linking them to the surface and through a number of nonthermal processes (global heating of the grains or sputtering by cosmic rays, chemical desorption). Species from the bulk can also be released in the gas phase by cosmic-ray sputtering. All the equations and details on the model can be found in \citet{2016MNRAS.459.3756R} and \citet{2021A&A...652A..63W} \citep[see also][]{2024A&A...689A..63W}. As usually assumed, only one grain size made of silicates of 0.1$\mu$m is considered. The gas to dust mass ratio is 100. \\

\subsection{Non-diffusive chemistry}\label{non-diffusive-section}

To this version of Nautilus, we added the non-diffusive chemistry using two different methods. The \citet{2020ApJS..249...26J} formalism is computed by taking into account the rates of production of species by other processes. The rate of the reaction (in s$^{-1}$) between species A and B on the grains is computed via
\begin{equation}
    \rm R_{AB} = \frac{f}{N_S} \left( R_{comp}(A)N(B) + R_{comp(B)}N(A)\right)
,\end{equation}
where f is the probability of the reaction to occur assuming the competition between diffusion and reaction. The term N$_{\rm S}$ is the number of sites on one grain, while N(B) and N(A) are the number of species A and B on one grain. The expression of $\rm R_{comp}$ is more complicated. It is the rate of appearance of species A or B. 

There is another formalism proposed by \citet{2018PCCP...20.5359S} where the rate of reaction is the diffusive rate \citep{1992ApJS...82..167H} and where the diffusion part is removed. Compared to the equation from \citet{2018PCCP...20.5359S}, we slightly changed the expression to take into account the dilution of the species: 
\begin{equation}
  \rm R_{AB}
  =
  f\!\left(
    \frac{\nu_A+\nu_B}{\max\bigl(N_{ab},\,N_S\bigr)}
  \right)
  N(A)N(B)
,\end{equation}
where $\nu_{\rm A}$ and $\nu_{\rm B}$ are the vibrational frequencies of species A and B, and N$_{\rm ab}$ is the number of molecules in the bulk or the surface. 
In this work, we compare the two methods.\\

\subsection{Radiolysis}
Following \citet{2018ApJ...861...20S}, we included in Nautilus the radiolysis for all species. 
We considered the radiolysis processes 
\begin{align*}
\rm A  &\rm \leadsto A^+ + e^- \rightarrow A^* \rightarrow B^* + C^*, \\
\rm A  &\rm \leadsto A^* \rightarrow B + C,  \\ 
\rm A  &\rm \leadsto A^*, 
\end{align*}
where the species with an asterisk are suprathermal species, which can react with thermal species without activation barriers or quenched by collision with the dust mantle. 
The radiolysis reactions are from \citet{2025ApJS..277....8L}.
The network involving suprathermal species has been constructed based on the kida.uva.2024 grain network. 
For each reaction involving thermal species that has its suprathermal counterpart as a reactant, we duplicated this reaction, changed this reactant into its suprathermal counterpart, and removed the activation barrier. 
If both reactants have their suprathermal counterparts, we duplicated this reaction twice and modified only one species to be suprathermal in each duplication. 
These barrierless reactions, including both the diffusive and non-diffusive manner, are expected to be able to promote the formation of large molecules in the ice surface and bulk.
In this work, we test the effect of this chemistry.  \\

\subsection{Chemical desorption}\label{chem-des-section}

We included in Nautilus the chemical desorption efficiency following \citet{Fredon2021}.  Two other formalisms are already present in Nautilus from \citet{2007A&A...467.1103G} and \citet{2016A&A...585A..24M}  \citep[see also][]{2016A&A...585A..55C}. All three methods depend on the fraction of the released energy that is dissipated to the grain surface.  The first method is an adhoc formalism where the fraction of desorbing species (number of desorbing species when a single reaction takes place) is expressed as
\begin{equation}
    \rm frac = \frac{aP}{1+aP}
.\end{equation}
Here, a is the fraction of energy remaining in the product and P is $\rm (1-E_D/E_{reac})^{s-1}$. In the latter, $\rm E_D$ is the binding energy of the product, $\rm E_{reac}$ is the enthalpy of formation of the species, and s is 2 for diatomic molecules and 3N-5 for more complex species (N being the number of atoms in the molecule). The parameter a is adjustable and suggested to be 0.01. For most if not all reactions, P is rather small so that frac is equal to a. 

The second method \citep{2016A&A...585A..55C} is based on experiments but only for very few and simple systems. The fraction of desorbing species is computed as\begin{equation}
    \rm frac = \exp{-\frac{E_D}{\epsilon E_{reac}/N}}
,\end{equation}
with N as the degree of freedom of the product. The term  $\epsilon = \frac{(M-n)^2}{(M+m)^2}$, where m is the mass of the product and M is the effective mass of the surface, which is an unknown parameter for the water ices. \citet{2016A&A...585A..24M} recommended using the M parameter of bare grains (120 amu) and divided the fraction by a factor of ten based on the experimental results. 

The third method is from \citet{Fredon2021}, and it is based on a comparison between rate equation models and Monte Carlo simulations. They defined the fraction of desorbing molecules as
\begin{equation}
   \rm frac = \theta \left( 1 - exp\left(-\frac{\chi \Delta E_{react} - E_D}{3E_D}\right) \right)
,\end{equation}
with $\chi$ as an empirical factor depending on the number of products (one or two). Note that for the two previous methods, only channels producing one species are considered, the other channels are set to zero desorption. The term f an empirical factor linked to where the species are on the surfaces. $\chi$ for two-produced species does not influence the results. For reactions producing only one species, we used the experimental results from \citet{2018NatAs...2..228O} to constrain $\chi$. In the experiments, they found that  H$_2$S formation via an H + HS reaction leads to a desorption of 60\% of the product. Inverting Fredon's formula, we found an $\chi$ parameter of 0.08. We use this value in this work and make a comparison to the other two methods. This final formalism has been included in our standard model (see below). These are the three formalisms we tested.

\subsection{Standard model and physical parameters}\label{standard-model}

\begin{table}
\caption{Elemental abundances.}
\label{ab-table}
\centering
\begin{tabular}{lccc}
\hline
\hline
Species & Abundance & Reference \\
\hline
He & $9\times 10^{-2}$ & 1 \\
N & $6.2\times 10^{-5}$ & 2 \\
O & $2.4\times 10^{-4}$ & 3 \\
C$^+$ & $1.7\times 10^{-4}$ & 2 \\      
S$^+$ & $8\times 10^{-8}$ &    4\\
Si$^+$ & $8\times 10^{-9}$ &  4\\
Fe$^+$ & $3\times 10^{-9}$ &  4\\
Na$^+$ & $2\times 10^{-9}$ & 4\\
Mg$^+$ & $7\times 10^{-9}$ &  4\\
P$^+$ & $2\times 10^{-10}$ & 4\\
Cl$^+$ & $1\times 10^{-9}$ &  4\\
F  & $6.68\times 10^{-9}$ &  5\\
\hline
\end{tabular}
\tablefoot{(1) See the discussion in \citet{2008ApJ...680..371W}. \\
(2) \citet{2009ApJ...700.1299J}. \\
(3) See the discussion in \citet{2011A&A...530A..61H}. \\
(4) Low metal abundance from \citet{1982ApJS...48..321G}.\\
(5) Depleted value from \citet{2005ApJ...628..260N}.}
\end{table}

\begin{table*}
\caption{Summary of the models.}
\label{models-table}
\begin{tabular}{ll}
\hline
\hline
Model & Description \\
\hline
Standard & without non-diffusive chemistry, \citet{Fredon2021} chemical desorption\\
Model 1 & with non-diffusive chemistry from \citet{2020ApJS..249...26J}, \citet{Fredon2021} chemical desorption\\
Model 2 & with non-diffusive chemistry from \citet{2018PCCP...20.5359S}, \citet{Fredon2021} chemical desorption\\
Model 3 & without non-diffusive chemistry, Minissale et al. (2016) chemical desorption\\
Model 4 & without non-diffusive chemistry, Garrod et al. (2006) chemical desorption\\
\hline
\end{tabular}
\end{table*}

Our standard model is the version in Nautilus not including non-diffusive chemistry nor radiolysis and using the chemical desorption of \citet{Fredon2021}. To simulate a cold core, we assumed the simple model in which the gas and grain temperature and H density are constant with time to 10~K and $2\times 10^4$~cm$^{-3}$, respectively. The visual extinction is high (20) in simulated shielded regions. The cosmic-ray ionization rate $\zeta$ is $1\times 10^{-16}$~s$^{-1}$. Note that we used a higher $\zeta$ compared to the typical values to get closer to the value found by \citet{2019A&A...624A.105F}. The initial conditions are atomic (neutral or ionic), except for H$_2$, which is molecular. The initial abundances are given in Table \ref{ab-table}. All the abundances presented in this paper are with respect to the proton density.

\section{Testing the various assumptions}\label{results-section}

We ran various models to test the processes described in Sects. \ref{non-diffusive-section} to \ref{chem-des-section}. We found no influence of the radiolysis itself without non-diffusive chemistry. The suprathermal species have such a low abundance and the dissociation rates are too low compared to the one induced by UV photons that reacting through normal diffusion mechanisms is not efficient. We therefore do not show any results of such models. 

Our standard model is the one described in Sect. \ref{standard-model}. Model 1 is our standard model with the non-diffusive chemistry by \citet{2020ApJS..249...26J}. Model 2 is the same but with the non-diffusive chemistry by \citet{2018PCCP...20.5359S}. Model 3 and 4 do not have any non-diffusive chemistry but the chemical desorption from \citet{2016A&A...585A..24M} and \citet{2007A&A...467.1103G} , respectively.  A summary of the models is given in Table \ref{models-table}.

\begin{figure*}
\centering
\includegraphics[width=1\hsize]{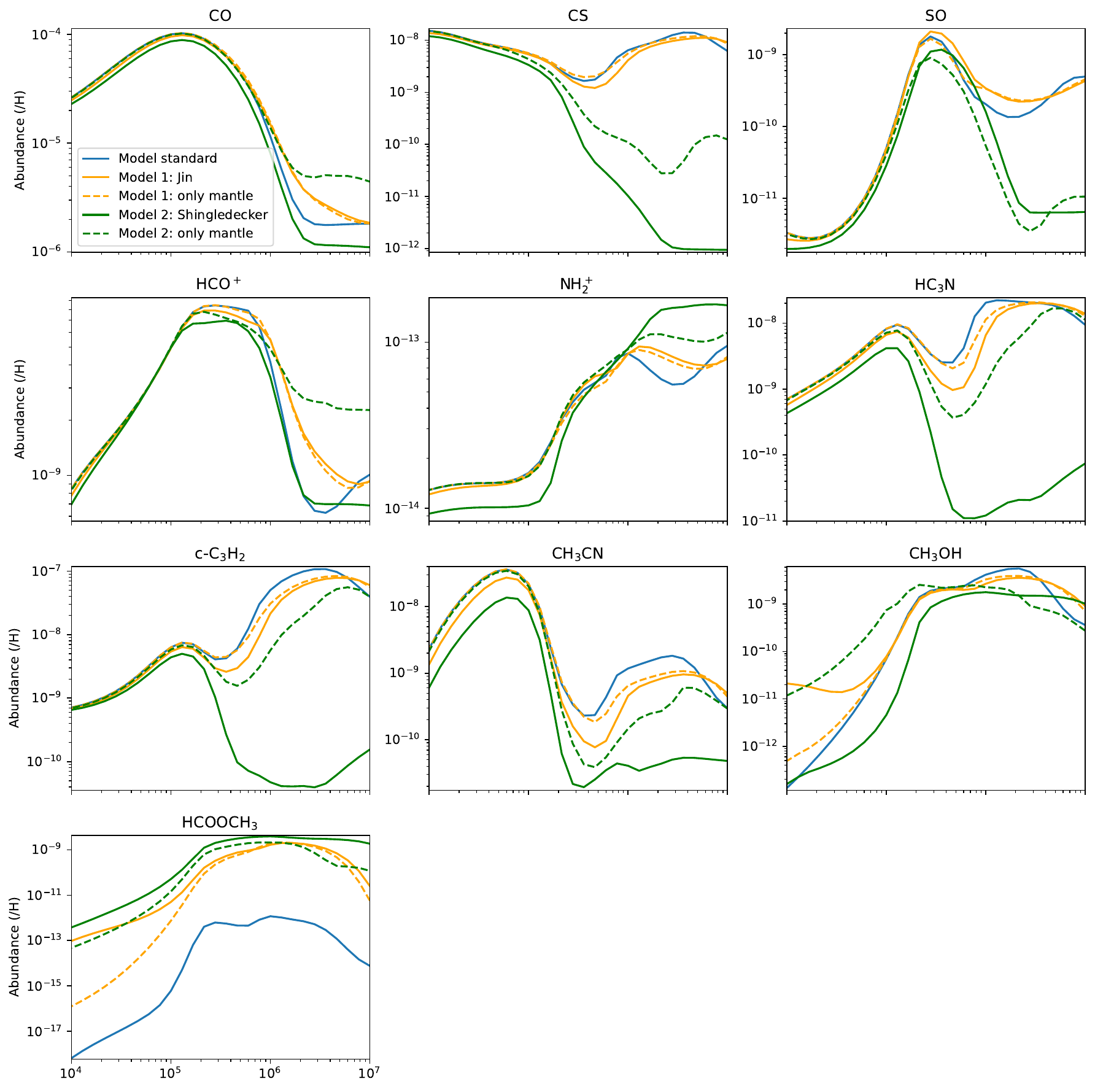}
\caption{Abundances as a function of time using our standard model (without non-diffusive chemistry), Model 1, and Model 2 for a selection of gas-phase species.
}
\label{stand-mod1-mod2}
\end{figure*}
\begin{figure*}
\centering
\includegraphics[width=1\hsize]{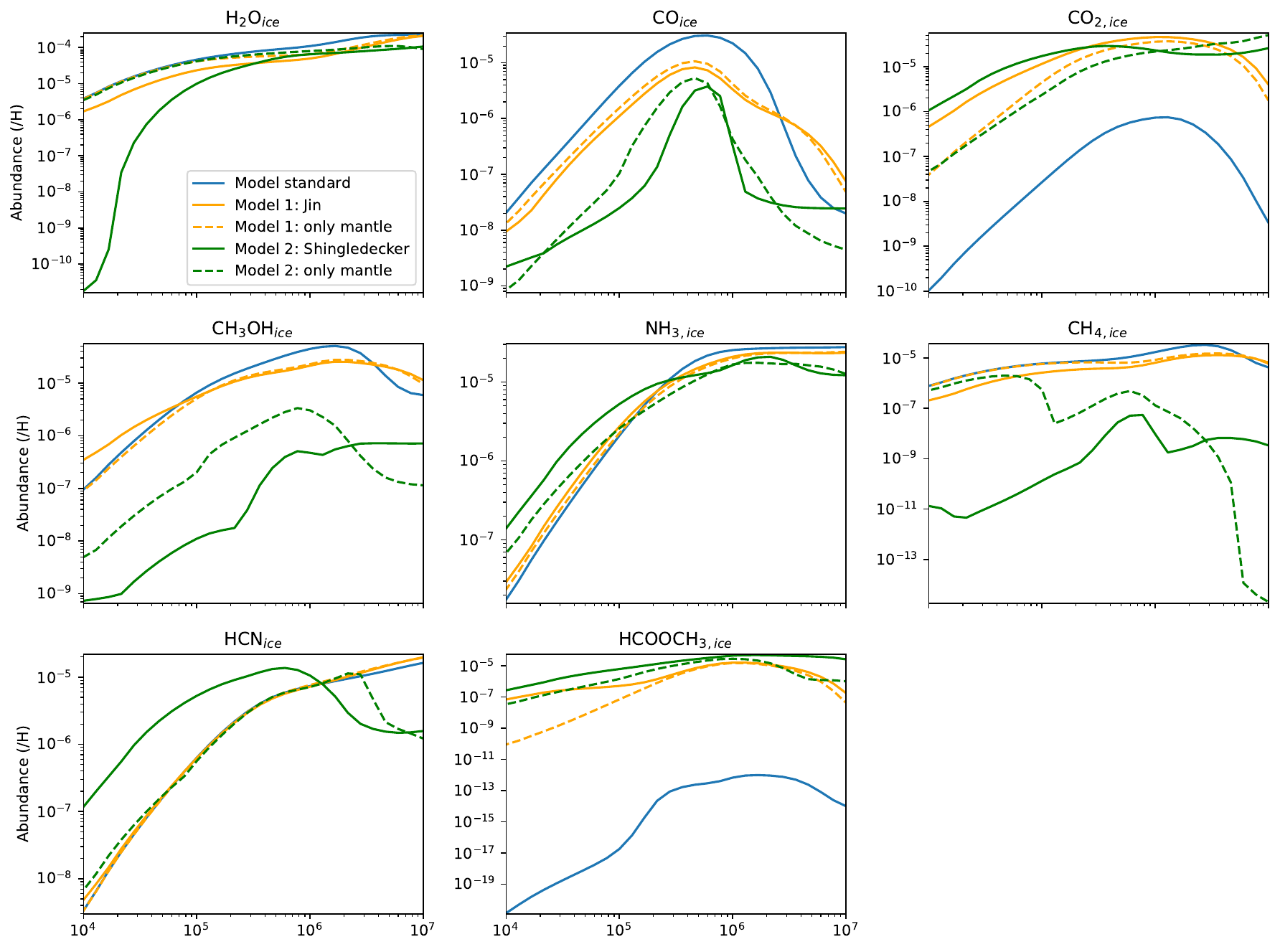}
\caption{Abundances as a function of time using our standard model (without non-diffusive chemistry), Model 1, and Model 2 for the main ice constituents.
}
\label{stand-mod1-mod2-ices}
\end{figure*}

\begin{figure*}
\centering
\includegraphics[width=1\hsize]{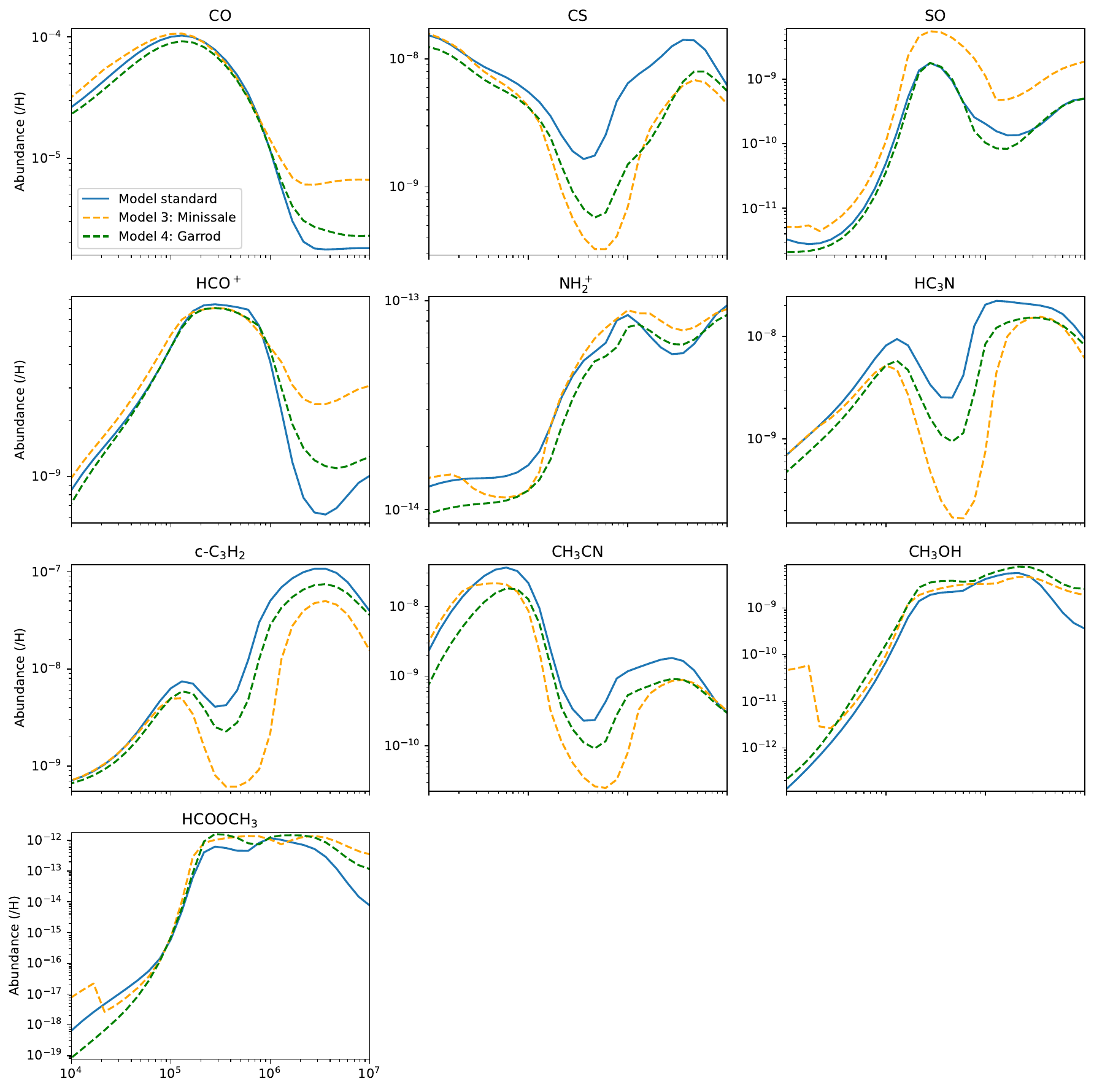}
\caption{Abundances as a function of time using our standard model (without non-diffusive chemistry), Model 3, and Model 4 for a selection of gas-phase species.
}
\label{stand-stand-mod3-mod4}
\end{figure*}
\begin{figure*}
\centering
\includegraphics[width=1\hsize]{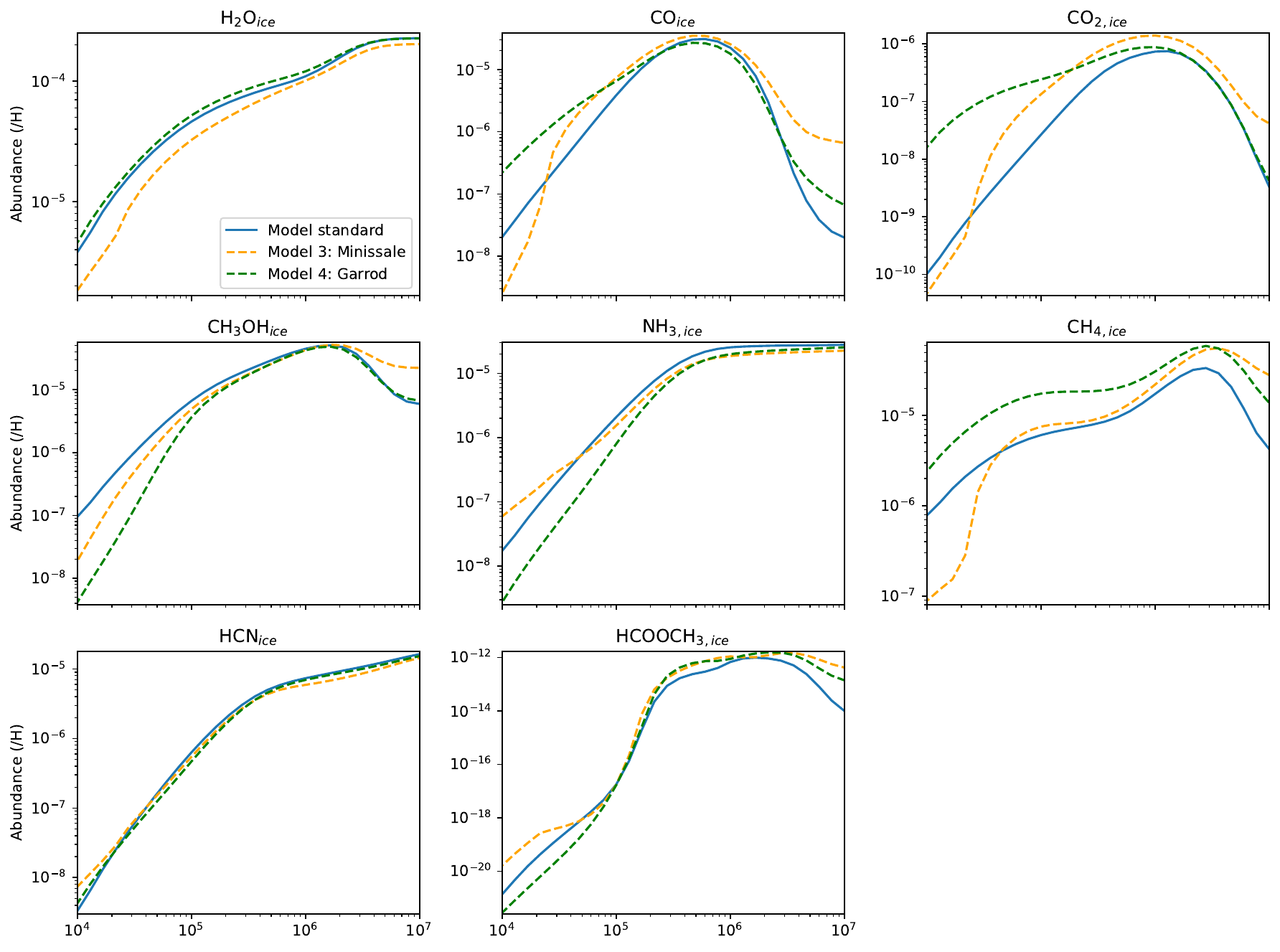}
\caption{Abundances as a function of time using our standard model (without non-diffusive chemistry), Model 3, and Model 4 for the main ice constituents.
}
\label{stand-stand-mod3-mod4-ices}
\end{figure*}

\subsection{Effect of the non-diffusive chemistry}

In the following, we first compare our standard model (without non-diffusive chemistry) with Model 1 (non-diffusive chemistry by Jin \& Garrod) and then Model 1 with Model 2 (non-diffusive chemistry by Shingledecker \& Herbst). We focus on a number of gas-phase and ice molecules commonly observed in cold cores and then discuss the ice composition in general.
The modeling results are shown in Figs. \ref{stand-mod1-mod2} and \ref{stand-mod1-mod2-ices}.

\subsubsection{Gas-phase abundances}

 

For the gas-phase species and the standard model compared to Model 1, the molecular abundances are not that different. The main difference is only a factor of a few. One exception is the complex organic molecule HCOOCH$_3$, which is enhanced by the non-diffusive chemistry by several orders of magnitude. 

Comparing Models 1 and 2 (the two formalisms for non-diffusive chemistry), most molecules exhibit a strong difference of several orders of magnitude at the typical age of cold cores ($10^6$~yrs), except for CO, HCO$^+$, N$_2$H$^+$, and CH$_3$OH. In nearly all cases, but N$_2$H$^+$, Model 1 produces larger abundances. From a general point of view, in Model 2, the main reservoirs of species are shifted toward species that necessitate heavy precursors to diffuse (see also Sect. \ref{reservoirs}). The CS and SO abundances predicted by Model 2 are much smaller than the observed abundances (see Fig \ref{all-obs-ab}). Note that we used an initial sulfur abundance of $8\times 10^{-8}$, which is typically used in chemical models, but if we increase it to its cosmic value, the CS and SO abundances become closer to the observed values. The HCN$_{\rm ice}$ becomes a reservoir of nitrogen in Model 2, as abundant as NH$_{\rm 3ice}$. Carbon is mainly in HCOOCH$_{\rm 3ice}$ and is more abundant than CO$_{\rm 2ice}$ (see also Sect. \ref{reservoirs}). 
Note that both CH$_3$OCH$_3$ (not shown on the figure) and HCOOCH$_3$ are much more abundant in the gas-phase when using Model 2. 
The gas-phase molecules are produced by the chemical desorption at this density \citep[see also][]{2021A&A...652A..63W}. 

In Models 1 and 2, all molecules from the surface and the mantle can undergo non-diffusive chemistry. In Fig. \ref{stand-mod1-mod2}, we overplot the abundances for these models with the non-diffusive chemistry only in the mantle. For Model 1, the results are only slightly changed. For Model 2, the impact is much stronger for CO, CS, HC$_3$N, c-CH$_3$H$_2$, and CH$_3$CN. Overall, the non-diffusive chemistry seems less efficient, and the molecular abundances are much higher if the non-diffusive chemistry is allowed only in the mantle. This also shows that the chemistry on the surface strongly influences the gas-phase molecular abundances.

\subsubsection{Ice abundances}

We compared our standard model with Model 1 for the ices, and the main differences are for CO$_{\rm 2ice}$ and HCOOCH$_{\rm 3ice}$, which are produced much more by non-diffusive chemistry (see Fig.~\ref{stand-mod1-mod2-ices}). The main difference between the model with and without non-diffusive chemistry from Jin \& Garrod (Model 1) concerns the formation of heavy molecules that are not formed by simple hydrogenation as expected. 
The abundance of other hydrogenated species, such as CH$_{\rm 4ice}$ and H$_2$O$_{\rm ice}$, and the one of CO$\rm _{ice}$ are decreased in the non-diffusive model because HCOOCH$_{\rm 3ice}$ takes a significant fraction of the carbon (see also Sect. \ref{reservoirs}). 

In Table~\ref{models-diff} we list the molecules that present the largest difference between our standard model and Model 1 at $10^6$ yr. The difference was computed as $\rm \log(X_{standard~model}) - \log(X_{Model~1})$, which means that a difference of -2 in the abundance in Model 1 is two orders of magnitude higher than in the abundance with Model 1. The minimal difference considered in the table is two orders of magnitude. All of these species are produced more with the non-diffusive chemistry, as they require heavy precursors to form. For instance, all sulfur-bearing species require sulfur to diffuse, which has a heavy binding energy of 2600~K \citep{2017MolAs...6...22W}. Some of these species are enhanced by 29 orders of magnitude or more. Among these six species, two have an abundance of about $10^{-9}$ at $10^{6}$ yr (H$_2$S$_{\rm 3ice}$ and NHCHO$_{\rm ice}$) with Model 1. NH$_2$CH$_2$SH$_{\rm ice}$ has an abundance of $10^{-11}$, while NH$_2$CHS$_{\rm ice}$ and NH$_2$CH$_{\rm 2ice}$ have an abundance of about $10^{-13}$. Last, S$_{\rm 3ice}$ has a very small abundance of $10^{-19}$. Among the species without sulfur, we found a number of complex organic molecules. HCOOCH$_{\rm 3ice}$ shows an increase of more than seven orders of magnitude, starting from an abundance of $10^{-12}$ with the standard model to an abundance of more than $10^{-5}$ with Model 1. This molecule is formed by the non-diffusive chemistry between HCO$_{\rm ice}$ and CH$_3$O$_{\rm ice}$. CH$_3$OCH$_{\rm 3ice}$ starts by the non-diffusive reaction between O$_{\rm ice}$ and C$_{\rm 2ice}$ to form CCO$_{\rm ice}$, followed by the successive hydrogenation. It has a high abundance of $3\times 10^{-6}$. CH$_3$OCH$_{\rm 2ice}$, CH$_3$CHO$_{\rm ice}$, H$_2$CCO$_{\rm ice}$, and CH$_3$CO$_{\rm ice}$ are formed during this process. HOCO$_{\rm ice}$ and HCOOH$_{\rm ice}$ are also linked starting from the non-diffusive reaction between OH$_{\rm ice}$ and CO$_{\rm ice}$. 

In Fig. \ref{stand-mod1-mod2-ices} we also plot the ice composition if the non-diffusive chemistry is only allowed in the mantle. In contrast to the gas-phase molecules, the effect here is less obvious. The main effect is on CO$_{\rm ice}$, CO$_{\rm 2ice}$, CH$_3$OH$_{\rm ice}$, CH$_{\rm 4ice}$, and HCOOCH$_{\rm 3ice}$. The main effect is a decrease of these molecule abundances at late times ($> 3\times 10^6$~yr). The reason for this decrease is a less efficient formation of these species in favor of CO$_{\rm 2ice}$.

\begin{table}
\caption{Difference between the standard model and Model 1. }
\label{models-diff}
\begin{tabular}{llll}
\hline
\hline
Molecule & diff & Molecule & diff\\
\hline
NH$_2$CH$_2$SH$_{\rm ice}$ &  -29.8 & HOCO$_{\rm ice}$ &  -4.6\\
H$_2$S$_{\rm 3ice}$  & -29.2 &  CH$_3$OCH$_{\rm 3ice}$ &  -4.0\\
NH$_2$CHS$_{\rm ice}$ &  -29.0 & HCOOH$_{\rm ice}$ &  -3.7\\
NH$_2$CH$_2$S$_{\rm ice}$ &  -29.0 &  CH$_3$OCH$_{\rm 2ice}$ & -3.2\\
S$_{\rm 3ice}$  & -21.7 &  CH$_3$CHO$_{\rm ice}$ &  -3.2\\
NHCHO$_{\rm ice}$ &  -20.0 &  CS$_{\rm 2ice}$ &  -3.0\\
S$_{\rm 4ice}$ &  -19.8 &  CH$_3$NCO$_{\rm ice}$ &  -2.6\\
S$_{\rm 5ice}$ &  -17.8 &  HSSH$_{\rm ice}$ &  -2.5\\
 S$_{\rm 6ice}$ &  -15.7 &  HSS$_{\rm ice}$ &  -2.3\\
 S$_{\rm 7ice}$ &  -13.7 &  H$_2$CCO$_{\rm ice}$ &  -2.1\\
 S$_{\rm 8ice}$ &  -11.7 &  H$_2$C$_3$O$_{\rm ice}$ &  -2.1\\
 HCOOCH$_{\rm 3ice}$ &  -7.2 &  CH$_3$CO$_{\rm ice}$ &  -2.0\\
 CH$_3$CH$_2$OH$_{\rm ice}$ &  -5.7 & & \\
\hline
\end{tabular}
\end{table}

Comparing Models 1 and 2 in Fig.~\ref{stand-mod1-mod2-ices}, except for water and NH$_3$, all species abundances are changed at $10^6$ yrs. CO$_{\rm ice}$, CH$_3$OH$_{\rm ice}$, and CH$_4{\rm ice}$ are produced much less in Model 2. CO$_2{\rm ice}$ is less decreased but still less abundant in Model 2, while HCN$_{\rm ice}$ and HCOOCH$_3{\rm ice}$ are  produced more and become reservoirs of N and C. In this model, HCN$_{\rm ice}$  is produced by the non-diffusive reaction between H$_{\rm ice}$ and CN$_{\rm ice}$, while CN$_{\rm ice}$ is enhanced in Model 2 by the UV induced photo-dissociation of HOCN$_{\rm ice}$ and HOCN$_{\rm ice}$ by the non-diffusive reaction between N$_{\rm ice}$ and HCO$_{\rm ice}$. HCOOCH$_3{\rm ice}$ is formed by the non-diffusive reaction between HCO$_{\rm ice}$ and $\rm CH_3O_{ice}$. 

Regarding the ice composition in general, we list in Table~\ref{models1and2-diff} the difference ($\rm \log(X_{Model~2}) - \log(X_{Model~1})$) between molecular abundances up to $10^{12}$ orders of magnitude. For all of these species, Model 1 predicts higher abundances than Model 2. The differences between the two models is extremely high for many species. 
Interestingly, the atomic abundance of S, C, and O are strongly decreased in Model 2. The abundance of many intermediate radicals, such as CH$_3{\rm ice}$, CH$_2{\rm ice}$, CH$_{\rm ice}$, NH$_2{\rm ice}$, and NH$_{\rm ice}$, is also decreased in Model 2. This is due to a more efficient molecular formation in Model 2.

\begin{table}
\caption{Difference between Models 1 and 2. }
\label{models1and2-diff}
\begin{tabular}{llll}
\hline
\hline
Molecule & diff & Molecule & diff\\
\hline
S$_{\rm ice}$  & -20.1 &  CH$_{\rm 2ice}$ &  -12.8\\
C$_{\rm ice}$  & -16.8 &  CH$_{\rm ice}$  & -12.7\\
 HCS$_{\rm ice}$ &  -15.0 & C$_6$H$_{\rm 3ice}$  & -12.7\\
 S$_{\rm 3ice}$  & -13.8 & O$_{\rm ice}$  & -12.6\\
 CH$_{\rm 3ice}$ &  -13.6 & NH$_{\rm 2ice}$ &  -12.5\\
 OH$_{\rm ice}$  & -13.3 & C$_3$H$_{\rm 5ice}$  & -12.4\\
 C$_4$H$_{\rm 3ice}$ &  -13.3 & NH$_2$CH$_2$S$_{\rm ice}$  & -12.3\\
 CH$_2$SH$_{\rm ice}$ &  -13.1 & NH$_{\rm ice}$  & -12.1\\
 C$_8$H$_{\rm 3ice}$ &  -13.0 & CH$_2$CCH$_{\rm ice}$  & -12.1\\
 S$_{\rm 4ice}$ &  -12.8 & & \\
 \hline
\end{tabular}
\end{table}



\subsubsection{Reservoirs}\label{reservoirs}

\begin{table*}
\caption{Main reservoirs for the standard model and Models 1 and 2.}
\label{reservoirs-table}
\centering
\begin{tabular}{lllllllll}
\hline
\hline
 Element & \multicolumn{2}{c}{O} & \multicolumn{2}{c}{C}& \multicolumn{2}{c}{N} & \multicolumn{2}{c}{S}\\
Model & Molecule & \% & Molecule & \% & Molecule & \% & Molecule & \%  \\
\hline
Standard & H$_2$O$_{\rm ice}$ & 41 & CH$_3$OH$_{\rm ice}$ & 26 & NH$_{\rm 3ice}$ & 40 & S & 31 \\ 
&CH$_3$OH$_{\rm ice}$ & 18 & H$_2$CO$_{\rm ice}$ & 22 & N$_{\rm 2ice}$ & 16 & HS$_{\rm ice}$ & 16 \\
& H$_2$CO$_{\rm ice}$ & 16 & CH$_{\rm 4ice}$ & 12 & HCN$_{\rm ice}$ & 12 & H$_2$S$_{\rm ice}$ & 16 \\
& CO$_{\rm ice}$ & 9 & CO$_{\rm ice}$ & 13 & CH$_3$NH$_{\rm 2ice}$ & 3 & NS$_{\rm ice}$ & 9\\
 & CO & 5 & HCN$_{\rm ice}$ & 4 & & & & \\
 \hline
 Model 1 & H$_2$O$_{\rm ice}$ & 20 & CO$_{\rm 2ice}$ & 26 & NH$_{\rm 3ice}$ & 32 & S & 65 \\
 & CO$_{\rm 2ice}$ & 37 & CH$_3$OH$_{\rm ice}$ & 12 & N$_{\rm 2ice}$ & 32 & CS & 5\\
 & HCOOCH$_{\rm 3ice}$ & 13 & HCOOCH$_{\rm 3ice}$ & 18 & HCN$_{\rm ice}$ & 11 & &\\ 
 & CH$_3$OH$_{\rm ice}$ & 8 & CO & 9 & CH$_3$NH$_{\rm 2ice}$ & 5 & & \\
 \hline
 Model 2 & HCOOCH$_{\rm 3ice}$ & 37 & HCOOCH$_{\rm 3ice}$ & 64 & NH$_{\rm 3ice}$ & 26 & H$_2$S$_{\rm 3ice}$ & 75 \\
 & H$_2$O$_{\rm ice}$ & 27 & CO$_{\rm 2ice}$ & 32 &  N$_{\rm 2ice}$ & 20 & S & 6 \\
 & CO$_{\rm 2ice}$ & 19 & HCN$_{\rm ice}$ & 7 & HCN$_{\rm ice}$ & 16 & & \\
 & CO & 6 & & & CH$_3$NH$_{\rm 2ice}$ & 9 & & \\
 \hline
\end{tabular}
\end{table*}

Table~\ref{reservoirs-table} presents the main reservoirs for the different families of molecules (O-, C-, N, and S-bearing species) at $10^6$~yr. The percentages represent the fraction of the molecules in each element times 100. 

The main reservoirs change from one model to the other, except for H$_2$O$_{\rm ice}$ and NH$_{\rm 3ice}$, which are the main reservoirs of oxygen and nitrogen, respectively. Most of the reservoirs are on the grains, except for CO, S, and CS. For O- and C-bearing species, one new species appears in Models 1 and 2, which is HCOOCH$_{\rm 3ice}$, as already stated in the previous sections. The H$_2$CO$_{\rm ice}$ molecule important in the standard model is not a reservoir in the two other models. For N-bearing species, we have the same molecules, NH$_{\rm 3ice}$ and N$_{\rm 2ice}$, as for the main ones. HCN$_{\rm ice}$ is also an important molecule for all models, although it has not been detected in ices \citep{2023NatAs...7..431M}. CH$_3$NH$_{\rm 2ice}$ is also expected to contain a significant amount of nitrogen in all models. Some elements are still expected to be abundant in the gas-phase despite the cloud age: CO and atomic sulfur for all models and CS for Model 1. 
Sulfur is expected to mostly be in the gas phase for the standard model and in Model 1, whereas it is expected to be in the second reservoir in Model 2. H$_2$S$_{\rm ice}$ is only predicted to be a reservoir in the standard model. In Model 2, the formation of H$_2$S$_{\rm 3ice}$ is extremely efficient. Even though it is not a reservoir in Model 1, its abundance is still high -- much higher than in the standard model. 
It should be noted that we do not claim that H$_2$S$_{\rm 3ice}$ is the reservoir of sulfur on grains because its chemistry is too uncertain. However, we can suggest that non-diffusive chemistry could produce heavy S-bearing molecules.  \citet{2020ApJ...888...52S} predicts that S$_{\rm 8ice}$ is the reservoir of sulfur instead of H$_2$S$_{\rm 3ice}$. When comparing our standard model with Model 2, the S$_{\rm 8ice}$ abundance is increased by more than two orders of magnitude in Model 2, but it still negligible. \citet{2020ApJ...888...52S} also allowed non-diffuse chemistry only for radicals and suprathermal species, while we used it for all reactions. However, this may not explain the difference, as it would probably decrease the efficiency of this process . 
The difference between the two predictions is probably due to different chemical networks. First, \citet{2020ApJ...888...52S} do not include H$_2$S$_{\rm 3ice}$. It includes, however, the reaction H$_2$S$_{\rm 2ice}$ + H$_{\rm ice}$ $\rightarrow$ HS$_{\rm ice}$ + H$_2$S$_{\rm ice}$, which probably efficiently destroys H$_2$S$_{\rm 2ice}$. In our model, this reaction is not included, and reactions  H$_2$S$_{\rm ice}$ + HS$_{\rm ice}$, H$_2$S$_{\rm 2ice}$ + S$_{\rm ice}$,  H$_2$S$_{\rm 2ice}$ + HS$_{\rm ice}$, and H$_{\rm 2ice}$ + HS$_{\rm ice}$ produce H$_2$S$_{\rm 3ice}$. Except for direct and indirect photo-dissociations, no reaction destroys H$_2$S$_{\rm 3ice}$, so it becomes a reservoir, preventing the formation of larger S$_{\rm n, ice}$. 

Laboratory experiments of ice irradiation seem to indicate that the final products of ice chemistry are spread among several products rather than in a single reservoir, as predicted by Model 2 for carbon and sulfur and by Model 1 for sulfur. For instance, for sulfur the irradiation of H$_2$S ices by \citet{2025NatCo..16.5571H} preferably produces H$_2$S$_{\rm n,ice}$, with n=2-8 and S$_{\rm 8ice}$ between 5 and 20\% of sulfur each. Note, however, that H$_2$S$_{\rm 2ice}$ is the major product in the experiment not predicted by our models. For carbon, \citet{2024A&A...687A.227M} irradiated a mixture of CO, H$_2$O, and CH$_3$OH ices. They found that the produced carbon species were distributed among a large number of molecules, including CO$_2$ and H$_2$CO, but also among a variety of complex organic molecules (including possibly HCOOCH$_3$). Our model predictions seem to be, qualitatively, not in strong disagreement with these experiments, but they tend to predict major reservoirs, while findings based on experiments tend to report that the elements are spread over a larger number of molecules. 

Interestingly, except for NH$_{\rm 3ice}$, all reservoirs of nitrogen are not detected or not detectable. \citet{2023NatAs...7..431M} lists three upper limits: 8.5\% for H$_2$S$_{\rm ice}$, 5\% for HCN$_{\rm ice}$, and  1\% for CH$_3$CN$_{\rm ice}$. The upper limit for the first molecule is not in agreement with the standard model. The upper limit for HCN$_{\rm ice}$ is only in agreement with the standard model. The upper limit on CH$_3$CN$_{\rm ice}$ is compatible with all models. \\

\subsection{Results comparing various chemical desorption formalisms}
\begin{figure}[htbp]
\centering
\includegraphics[width=0.95\hsize]{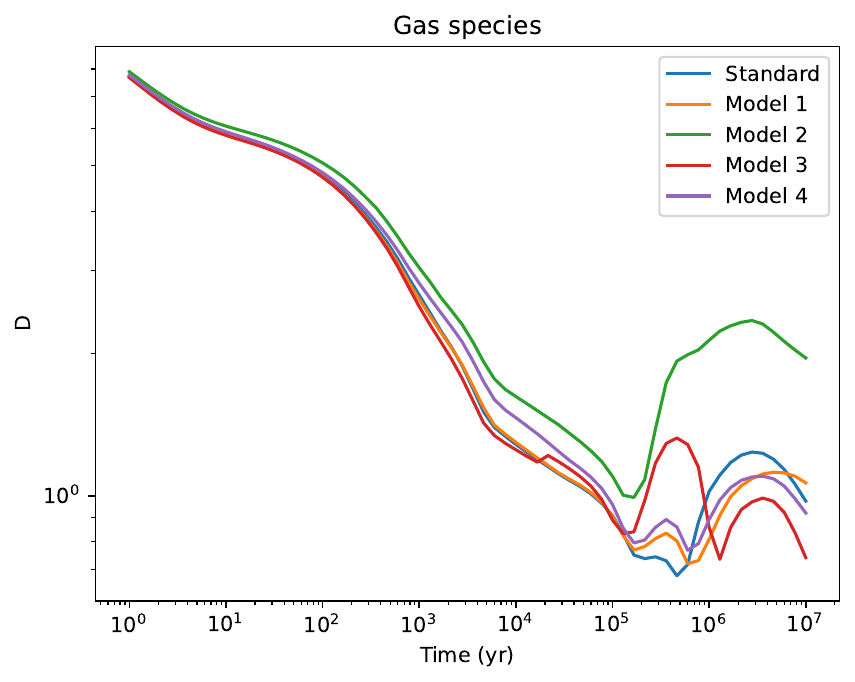}
\caption{Distance of disagreement for gas-phase molecules and all models as a function of time.
}
\label{D-allmods}
\end{figure}

\begin{figure*}[htbp]
\captionsetup{justification=raggedright,singlelinecheck=false}
\centering
\includegraphics[width=0.8\hsize]{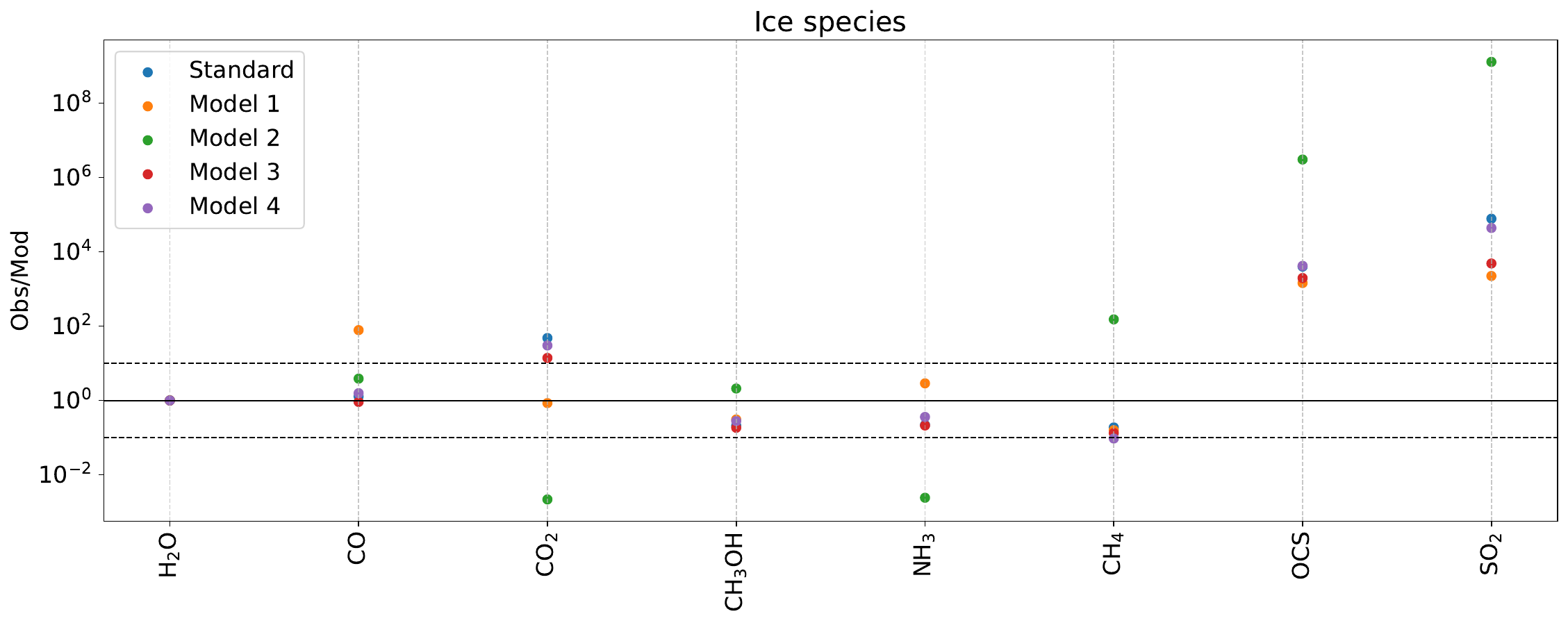}
\includegraphics[width=0.8\hsize]{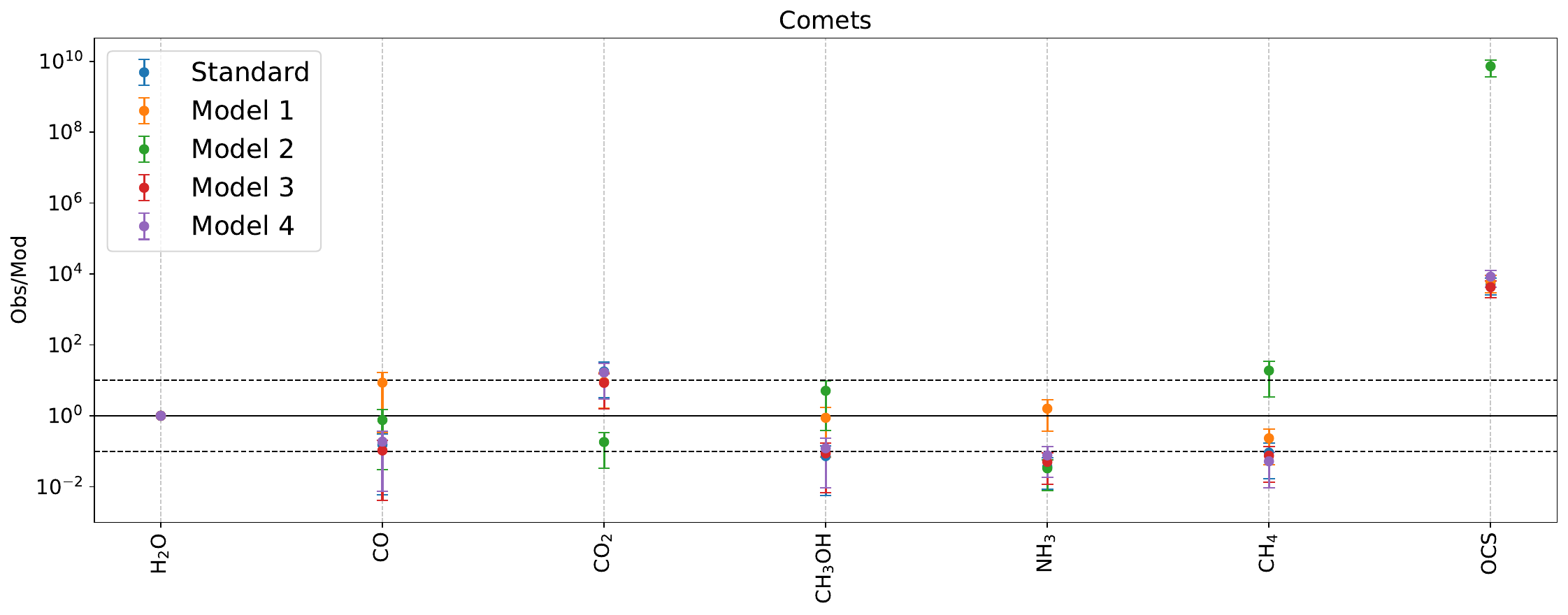}
\caption{Ratio between observed and modeled abundances for all ice species and in comets at the best D.
}
\label{comp-obs_ab-ice}
\end{figure*}
Using the same species as in the previous sections, we compared the model results of our standard model \citep[following][]{Fredon2021} with the two other chemical desorption formalisms: Model 3 for \citet{2016A&A...585A..24M}  and Model 4 for \citet{2007A&A...467.1103G}. Figs \ref{stand-stand-mod3-mod4} and \ref{stand-stand-mod3-mod4-ices} show the results for the three models. As expected, the ice composition is not influenced much by the formalism of chemical desorption compared to the gas-phase abundances. Interestingly, some species not formed on the grains, such as SO, CS, HC$_3$N, and c-C$_3$H$_2$, show differences up to one order of magnitude. 

CS is produced by the gas-phase reaction between H and HCS, and HCS by the reaction between C and H$_2$S. HCS is more abundant in the standard model because C is more abundant. At $10^6$~yr, C is produced by the abstraction reaction between H and CH. CH is produced on the grains by the reaction between C and H, which is followed by chemical desorption. Chemical desorption is more efficient in the standard model (frac = 0.28), with Model 4 following next in efficiency (frac=0.09) and then Model 3 (frac=0.08). 

SO is produced more in Model 3, while it is similar in the two other models. Tracing back its formation in the gas phase, it starts by the formation of H$_2$O in the ices followed by chemical desorption of the product. While the desorption is negligible for the standard model and Model 4, it is 24\% in Model 3. This last ratio is a direct result of the experiments conducted by \citet{2016A&A...585A..24M}. H$_2$O reacts in the gas phase with H$_3^+$ to form H$_3$O$^+$ and subsequently forms OH, and OH reacts with S to form SO.

From a general point of view, the higher abundances of the C molecules (HC$_3$N, c-C$_3$H$_2$, and CH$_3$CN) in the standard model traces back to the larger carbon abundance in the gas phase that is followed by a series of reactions in the gas phase or at the surface of the grains. HC$_3$N and CH$_3$CN chemistry are related to the CN molecule, which is more abundant in the standard model, with Model 4 having the second-highest abundance. c-C$_3$H$_2$ is formed by a series of ion-neutral reactions, which starts from C. CH$_3$CHO is also formed in the gas-phase through O + C$_2$H$_5$ $\rightarrow$ H + CH$_3$CHO. Again C$_2$H$_5$ is more abundant in the standard model and in Model 4. 

\section{Comparison with observations}\label{obs-section}

To test the various hypotheses presented above, we compared the model results to the observed abundances listed in Appendix \ref{obs}. To define the best agreement, we computed the difference between observed and modeled abundances as a function of time, also called the distance of disagreement D:
\begin{equation}
\rm D(t) = \frac{1}{N_i} \sum_i \left|log(X_{obs,i}-log(X_{mod,i})\right|
.\end{equation}

Figure \ref{D-allmods} shows the distance of disagreement for all models (standard to Model 4). Model 2 (with non-diffusive chemistry from \citealt{2018PCCP...20.5359S}) gives the poorest agreement among the models, while the standard model gives the best at $5\times 10^5$ yr. Model 3 and Model 1 also give good agreement, though at a slightly later point. This means that neither the non-diffusive chemistry in Model 2 nor the chemical desorption in Model 3 improves the agreement between the observation and the model. The best distance of disagreement D for the standard model is 0.7, which means that the mean difference between the observed and modeled abundances is a factor of five. Figure \ref{comp-obs-ab} shows the ratio between the observed and modeled abundances for the best times of all five models. The species whose abundance is larger than a factor of ten for all models are C$_3$O, C$_4$H$_2$, C$_6$H$_2$, CN, H$_2$CN, C$_5$N, HCS$^+$, and CCS.

We performed the same work for the ice species as observed by \citet{2023NatAs...7..431M} with JWST. The distance of disagreement values are shown in Fig. \ref{D-allmods-ice}. All models but Model 2 show a similar agreement for a similar time, with Models 1 and 3 being slightly better: D is 1.3 (meaning a mean difference by a factor of 20) and has a time of $4\times 10^5$ yr. Overall, the agreement is not very good, but Model 2 clearly provides the worst agreement with the observations. Figure \ref{comp-obs_ab-ice} (upper panel) shows the ratio between observed and modeled abundances for ice species at the best time of each model. The worst agreement for Model 2 comes from an underestimation of the abundance of CH$_{\rm 4ice}$, OCS$_{\rm ice}$, and SO$_{\rm 2ice}$, while CO$_{\rm 2ice}$ and NH$_{\rm 3ice}$ are overestimated. CO$_{\rm 2ice}$ are underestimated by the models, except in Model 1. As noted by \citet{2023A&A...675A.165C}, CO$_2$ formation needs the diffusion of CO or O in the surfaces and has to overcome an activation barrier. The non-diffusion chemistry in Model 1 is then efficient to form CO$_{\rm 2ice}$, while Model 2 is too efficient. NH$_{\rm 3ice}$ is too efficiently produced by Model 2, while CH$_{\rm 4ice}$, OCS$_{\rm ice}$, and SO$_{\rm 2ice}$ are not produced enough. Methanol, NH$_{\rm 3ice}$, and CH$_{\rm 4ice}$ are well reproduced by the other models. Using the higher elemental abundance of sulfur, the overall D values are slightly better for the standard model and Model 1. The agreement for OCS$_{\rm ice}$ is better but still highly underproduced by the models, while the CH$_{\rm 4ice}$ abundance is underproduced by Model 2.

We also compared our model to observations in comets. For this, we used the review paper by \citet{2011ARA&A..49..471M}. In Fig. \ref{D-allmods-comets} we plot the distance of disagreement for comets, while in Fig. \ref{comp-obs_ab-ice} (lower panel) we plot the ratio between modeled abundances with respect to water with the observed ones in comets. The vertical bars represent the variations among comets. 
The first striking result is that the overall agreement is very similar to interstellar ices but with a worse agreement with Model 2, while Model 1 has a slightly better agreement. As can be seen in Fig. \ref{comp-obs_ab-ice}, the main species not reproduced by the models in OCS, which are underestimated by the models. Using the higher sulfur elemental abundance does not improve the agreement for OCS, while it does not change much for the other species. 

The CO$_{\rm ice}$ over the water ratio is also overpredicted by all models as well as the CH$_{\rm 4ice}$ over the water ratio. Model 2 also underestimates OCS$_{\rm ice}$. SO$_{\rm 2ice}$ was not observed or detected in comets. This again shows that the ratio of the molecules over water is very similar in comets and interstellar ices.

\section{Summary and conclusions}

We have examined the differences of the model predictions produced by various assumptions on the radiolysis with and without non-diffusive chemistry, the diffusive chemistry itself, and the chemical desorption. Several formalism have been tested for the non-diffusive chemistry and chemical desorption. 
The following conclusions can be drawn from our work:
\begin{itemize}
    \item Radiolysis without non-diffusive chemistry has no impact on the model predictions. 
    \item Non-diffusive chemistry from \citet{2020ApJS..249...26J} has an effect of only a factor of a few for gas-phase observed molecules in cold cores when compared to our model without non-diffusive chemistry. Only heavy species, such as complex organic molecules, are produced more by the non-diffusive chemistry.
    \item When comparing the two formalisms of non-diffusive chemistry, we found strong differences. The one of \citet{2018PCCP...20.5359S} changes the main reservoirs of the species in the ices, impacting the species in the gas-phase as well. The main reservoirs become H$_2$S$_{\rm 3ice}$ for S and HCOOCH$_{\rm 3ice}$ for C. For N, both NH$_{\rm 3ice}$ and N$_{\rm 2ice}$ remain the main reservoirs in all models.
    \item Allowing non-diffusive chemistry in the mantle only significantly reduces the formation of molecules in Model 2 with the formalism of \citet{2018PCCP...20.5359S}. 
    \item Model 2 represents a very fast bulk chemistry  capturing qualitatively experimental results, but it predicts major reservoirs of carbon and sulfur, while the experiments show that the elements are spread over a large number of molecules. 
    \item From comparison of the three formalisms for chemical desorption, we found almost no differences for the ice species. The difference in the gas-phase can be up to a factor of ten.
    \item We also compared the results of our models with gas-phase and ice observations in cold cores. Our standard model \citep[without non-diffusive chemistry and the chemical desorption by][]{Fredon2021} gives the best agreement with the gas-phase observed abundances in TMC-1 CP within a factor of five between observed and modeled abundances at $5\times 10^5$ yr, while Model 3 \citep[no non-diffusive chemistry and chemical desorption by][]{2016A&A...585A..24M} and Model 1 (non-diffusive chemistry by Jin \& Garrod 2020 and chemical desorption by Fredon et al. 2021) also show a good agreement.  For ice species and species observed in comets, it was impossible to discriminate among Models 1, 3, and 4, while Model 2 strongly disagrees with the observations, especially for S-bearing species. Using a higher elemental abundance of sulfur does not improve the agreement.
    
\end{itemize}

\begin{acknowledgements}
T. Tu acknowledges the financial support of the China Scholarship Council (No. 202406190190). VW thanks the CNRS program ``Physique et Chimie du Milieu Interstellaire'' (PCMI) co-funded by the Centre National d’Etudes Spatiales (CNES). We thank the anonymous referee for their constructive remarks, which helped improve strongly the paper.
\end{acknowledgements}

\bibliographystyle{aa}
\bibliography{bib}

\begin{appendix}
\cleardoublepage

\section{Observed abundances}\label{obs}
\FloatBarrier
In Fig.~\ref{all-obs-ab} we display the observed abundances listed in \citet{2013ChRv..113.8710A} and those computed by \citet{2025ApJS..281....9X} in TMC-1 (CP) assuming a H$_2$ column density of $2\times 10^{22}$~cm$^{-2}$. Some molecules are present in both of them and, except for a small number of them, they give similar abundances. For such cases, we selected the abundance from \citet{2025ApJS..281....9X} because it is a self-consistent calculation while the other reference is a list of molecules obtained by various studies.

\begin{center}
\includegraphics[width=\textwidth]{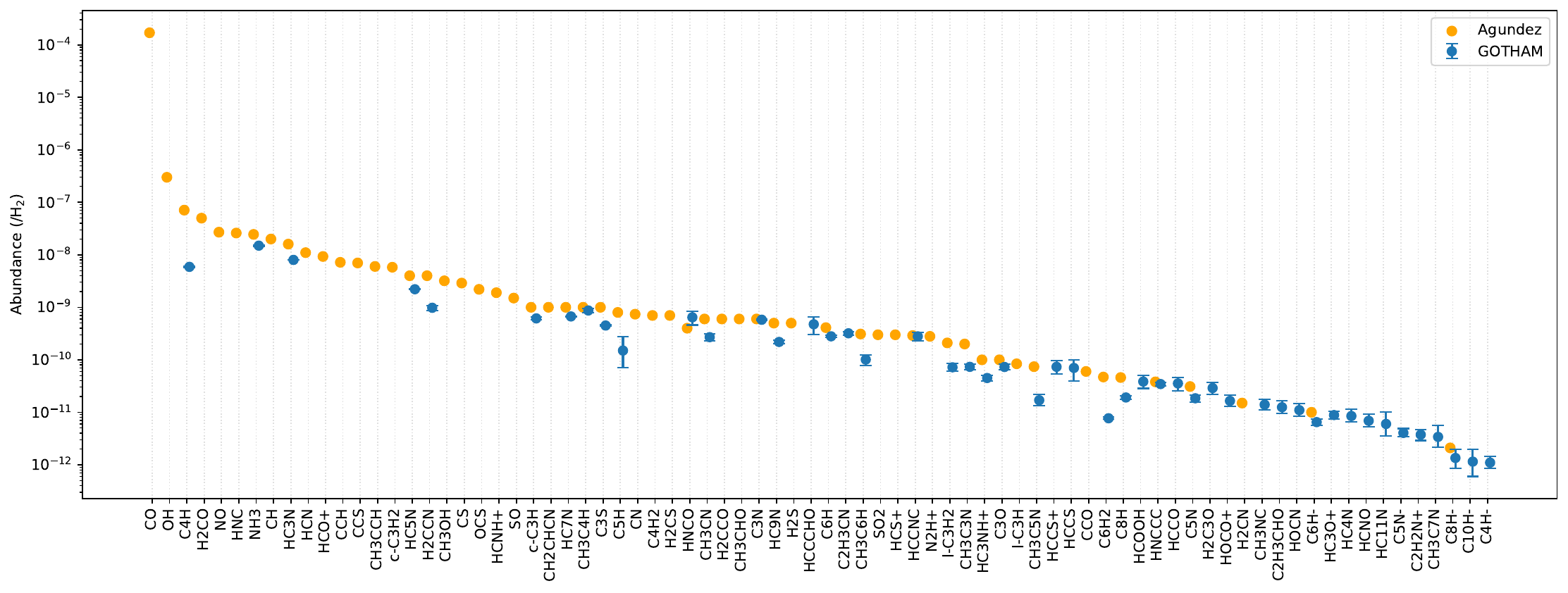}
\parbox{\textwidth}{
\captionof{figure}{Observed abundances in the gas-phase in TMC-1 (CP) listed in \citet{2013ChRv..113.8710A} and \citet{2025ApJS..281....9X}.}
\label{all-obs-ab}
}
\end{center}



\section{Comparison between observations and models}

\begin{figure}[htbp]
\centering
\includegraphics[width=1\hsize]{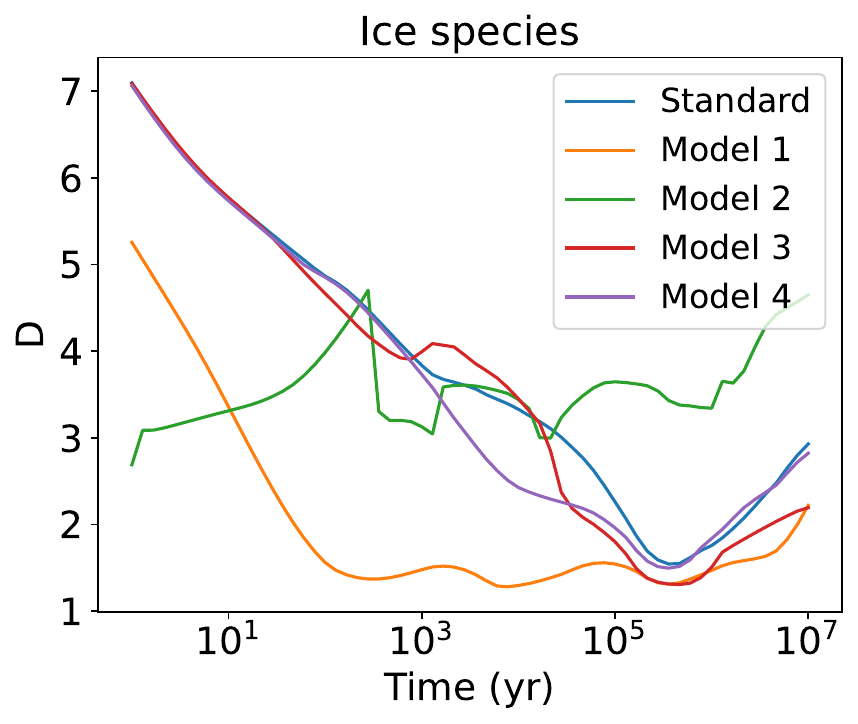}
\caption{Distance of disagreement for ice species. }
\label{D-allmods-ice}
\end{figure}

\begin{figure}[htbp]
\centering
\includegraphics[width=1\hsize]{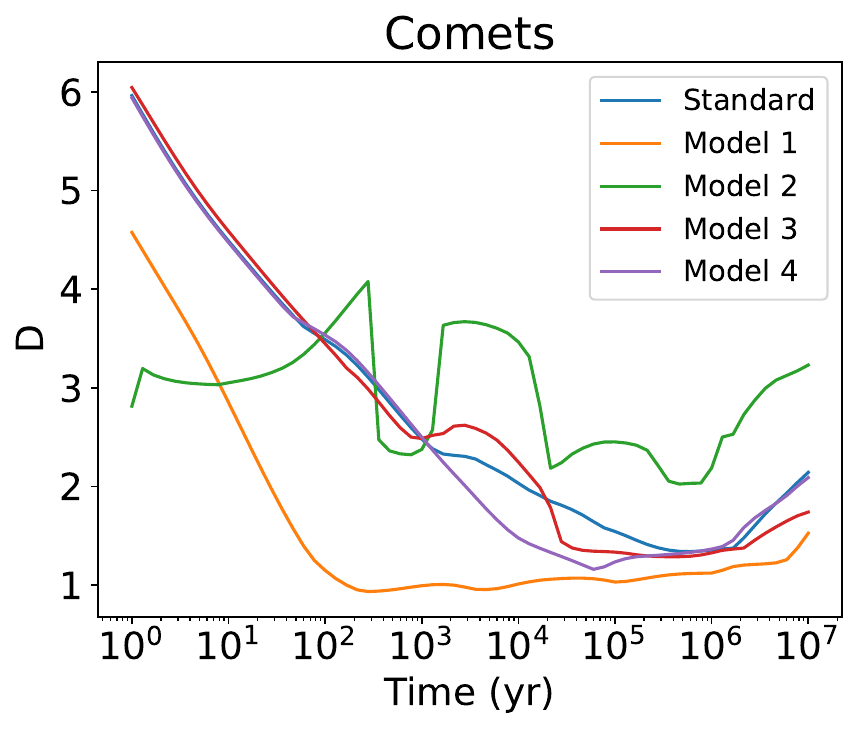}
\caption{Distance of disagreement for species observed in comets. }
\label{D-allmods-comets}
\end{figure}

Figure \ref{comp-obs-ab} shows the ratio between the observed abundances (see Appendix \ref{obs}) and the modeled abundances for all 5 models (see Table \ref{models-table}). The time of the best D is used as a reference to obtain the modeled abundance for each model. Figs. \ref{D-allmods-ice} and \ref{D-allmods-comets}  show the distance of disagreement for ice species and comets as a function of time. 

\begin{center}
\includegraphics[width=\textwidth]{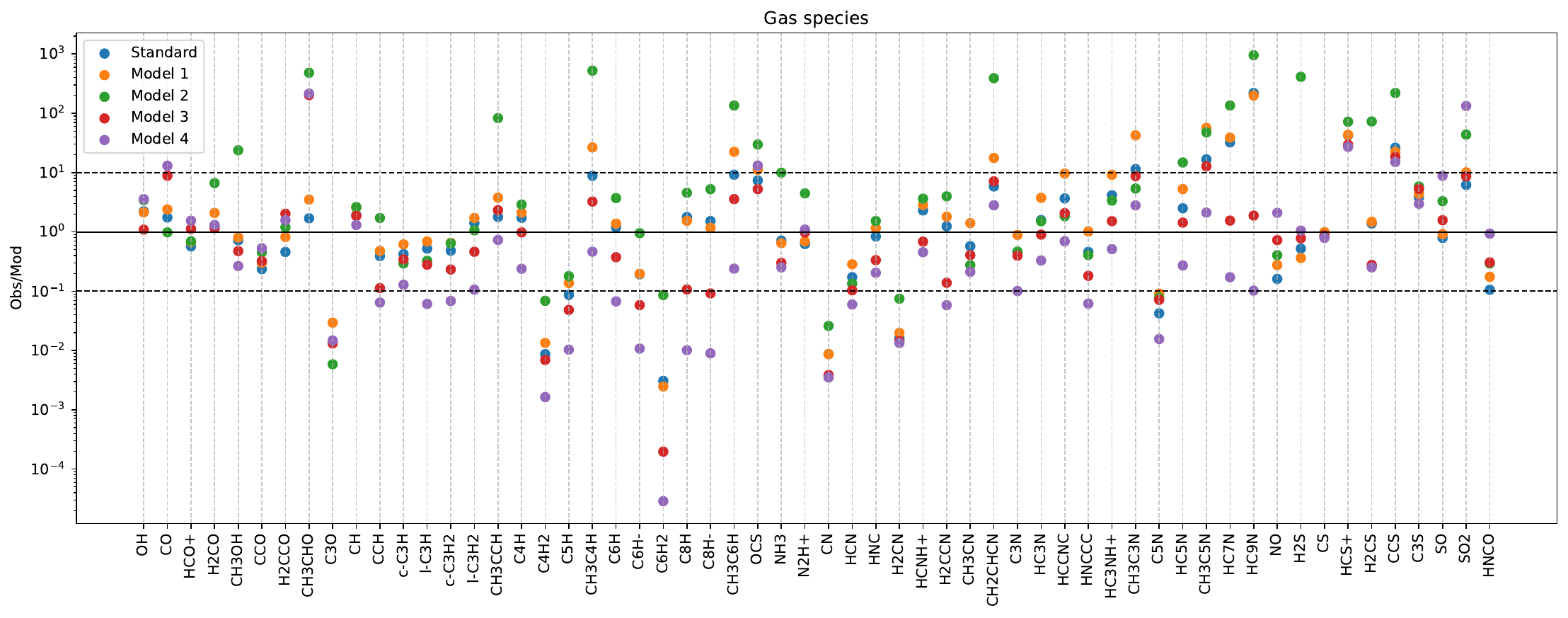}
\parbox{\textwidth}{
\captionof{figure}{Comparison between observed and modeled abundances for all observed molecules in the gas phase at the best time. }
\label{comp-obs-ab}
}
\end{center}


\clearpage

\end{appendix}
\end{document}